\documentclass[amsmath,amssymb,14pt]{article}

\usepackage{graphicx, epsfig}

\usepackage{amssymb,amsmath,xcolor}
\usepackage{amsthm}
\usepackage[a4paper, total={6.5in, 10in}]{geometry}

\usepackage{tikz}
\usetikzlibrary{patterns, quotes, positioning, decorations.markings}

\usetikzlibrary{arrows.meta,
                decorations.markings,
                patterns.meta}

\renewcommand\Re{\mathrm{Re}}
\renewcommand\Im{\mathrm{Im}}

\numberwithin{equation}{section}

\usepackage{subcaption}
\usepackage{caption}
\usepackage[export]{adjustbox}

\usepackage[nottoc]{tocbibind}

\usepackage{dcolumn}
\usepackage{bm}

\usetikzlibrary{calc, decorations.markings}
\usepackage{pgfplots}

\usetikzlibrary{external}

\newtheorem{mytheo*}{Theorem}
\newtheorem{mylemma*}{Lemma}
\newtheorem{myproposition*}{Proposition}
\newtheorem{mydef*}{Definition}[section]
\newtheorem{myremark*}{Remark}
\newtheorem{myproof*}{Proof}
\newtheorem*{mynotation*}{Notation}
\newtheorem{mycorollary*}{Corollary}
\newtheorem*{theorem*}{Theorem}

\usepackage{xcolor}

\DeclareMathOperator{\Real}{Re}

\DeclareMathOperator{\res}{Res}

\DeclareMathOperator{\sech}{sech}
\DeclareMathOperator{\sgn}{sgn}

\usepackage{mathtools}

\usepackage{authblk}
\usepackage{dsfont}

\begin{document}
\title{\textsc{Completeness of Energy Eigenfunctions for the Reflectionless Potential in Quantum Mechanics}}

\author[1]{Fatih Erman}
\author[2, 3]{O. Teoman Turgut}
\affil[1]{Department of Mathematics, \.{I}zmir Institute of Technology, Urla, 35430, \.{I}zmir, Turkey}
\affil[2]{Department of Physics, Bo\u{g}azi\c{c}i University, Bebek, 34342, \.{I}stanbul, Turkey}
\affil[3]{Department of Physics, Carnegie Mellon University, Pittsburgh, PA, United States}
\affil[1]{fatih.erman@gmail.com}
\affil[2]{turgutte@boun.edu.tr}

\maketitle

\begin{abstract} 
There are few exactly solvable potentials in quantum mechanics for which the completeness relation of the energy eigenstates can be explicitly verified. In this article, we give an elementary proof that the set of bound (discrete) states together with the scattering (continuum) states of the reflectionless potential form a complete set. We also review a direct and elegant derivation of the energy eigenstates with proper normalization by introducing an analog of the creation and annihilation operators of the harmonic oscillator problem. We further show that, in the case of a single bound state, the corresponding wave function can be found from the knowledge of continuum eigenstates of the system. Finally, completeness is shown by using the even/odd parity eigenstates of the Hamiltonian, which provides another explicit demonstration of a fundamental property of quantum mechanical Hamiltonians. 
\end{abstract}

\section{Introduction} \label{introduction}

The spectrum of a quantum Hamiltonian consists of the set of eigenvalues $E_n$  with finite degeneracies (or multiplicities) and the set of generalized eigenvalues $E$ (as continuous energy states). The eigenfunctions associated with $E_n$ are normalizable and satisfy $H \psi_n = E_n \psi_n$. However, the continuum eigenfunctions (so-called generalized eigenfunctions) $\psi_E$ corresponding to the continuous spectrum labeled by $E$ are not square integrable. The plane wave $(2\pi)^{-1/2} e^{ikx}$ is the well-known continuum
eigenfunction of the free particle Hamiltonian associated with the continuous spectrum labeled by $E=\hbar^2 k^2/2m$ in one dimension. These are indeed the eigenstates of the momentum operator and hence properly labeled by $k$ \footnote{These are indeed called the generalized eigenfunctions since they are not elements of $L^2(\mathbb{R})$ \cite{Faddeev}. In general, the proper meaning of these continuum eigenfunctions must be understood in the context of so-called Rigged Hilbert spaces or Gelfand triple (see e.g., Ref. \cite{Arnobook} and Ref. \cite{Appel} for the details) although our aim here is not to give a rigorous presentation.}. According to the spectral theorem, the set of energy eigenfunctions $\psi_n$ and $\psi_E$ together form a complete set of basis vectors so that one can expand an arbitrary square-integrable function in terms of them.  This is in fact the infinite dimensional generalization of the expansion of a vector with respect to a basis of eigenvectors of any Hermitian matrix in linear algebra. The full set of eigenfunctions is sufficient to describe the quantum system and this is one of the essential properties of quantum mechanics, known as the completeness relation of eigenfunctions \cite{Griffiths, Gasiorowicz}.

The completeness relation implies that a quantum mechanical system {\it must have} bound states if and only if the continuum eigenstates form an {\it incomplete set of vectors}. There are a few standard examples in which the completeness relation has been verified explicitly for systems having only discrete or only continuous spectra. The free particle and the single particle in a box are the most well-known examples in the standard textbooks of quantum mechanics \cite{Griffiths, Gasiorowicz}. The completeness relation for systems having both bound states and continuum states, such as the Dirac delta potential in one dimension \cite{Patil} and the Coulomb potential in three dimensions, \cite{Mukunda} have also been demonstrated explicitly by appropriate normalization of the eigenfunctions. When the lack of completeness of the continuum eigenstates corresponds to a one-dimensional projection, it may then be possible to calculate the bound state wave function from the completeness relation. Indeed this observation was first illustrated in Ref. \cite{Brownstein} by a concrete example in one dimension, where the potential is given by the Dirac delta function. This has been also discussed more recently in a quantum mechanics textbook \cite{Konishi}. However, no other explicitly solvable example has been worked out in the literature to the best of our knowledge.

The reflectionless potential well is another exactly solvable one-dimensional problem in quantum mechanics:
\begin{eqnarray}
V(x)= - \frac{\hbar^2 \kappa^2}{2m} \frac{N(N+1)}{\cosh^2 \kappa x} \;,
\end{eqnarray}
where $N$ is a positive integer and the parameter $\kappa>0$ is introduced for a dimensional reason. This potential for any positive number $N$ is also often referred to as the modified version of the original P\"{o}schl-Teller potential \cite{PoschlTeller} and the solution is available in several quantum textbooks \cite{Flugge, Landau}. The solution of the above reflectionless potential has been also discussed in the pedagogical paper \cite{Lekner} and as a textbook problem in \cite{Griffiths}. The family of potentials given above for any integer $N$ is reflectionless for an incoming wave of any energy because of the observation that there is no reflection part ($e^{-ikx}$) for an incoming state ($e^{ikx}$) \cite{Crandall, CrandallLitt}. An interesting realization of this potential was first proposed by Epstein \cite{Epstein} in the context of optics and later by Eckart \cite{Eckart} in a quantum mechanical problem. The bound state and scattering solutions of this potential can be expressed in terms of elementary functions by following the so-called factorization method \cite{Dong} or equivalently by the supersymmetric formulation of quantum mechanics \cite{CooperKhareSukhatme}, which is basically a generalization of the method used in solving the  harmonic oscillator potential by means of the creation and annihilation operators. What makes this form of the reflectionless potential interesting to study is its critical role in understanding the soliton (spatially localized wave with some additional properties) solutions of a non-linear wave equation, namely Korteweg-de-Vries (KdV) equation  in inverse scattering theory \cite{Novikov, DauxoisPeyrard}.

In this paper we will first review the solutions of the above particular form of the reflectionless potential by the factorization/susy algebra method. We will then verify the completeness relation by first normalizing the continuum eigenfunctions from their orthonormality condition, then showing that the continuum eigenstates satisfy the completeness relation only if we include the bound states as well. We also show that the same result can be obtained by considering the even and odd parity eigenstates of the Hamiltonian. Finally, we deduce the bound state wave function of the system only from the knowledge of the continuum eigenfunctions associated with a continuous spectrum.

\section{An Algebraic Method for Finding the Energy Eigenstates of the Reflectionless Potential}

\begin{figure}[h!]
    \centering
    \includegraphics[width=0.8\linewidth]{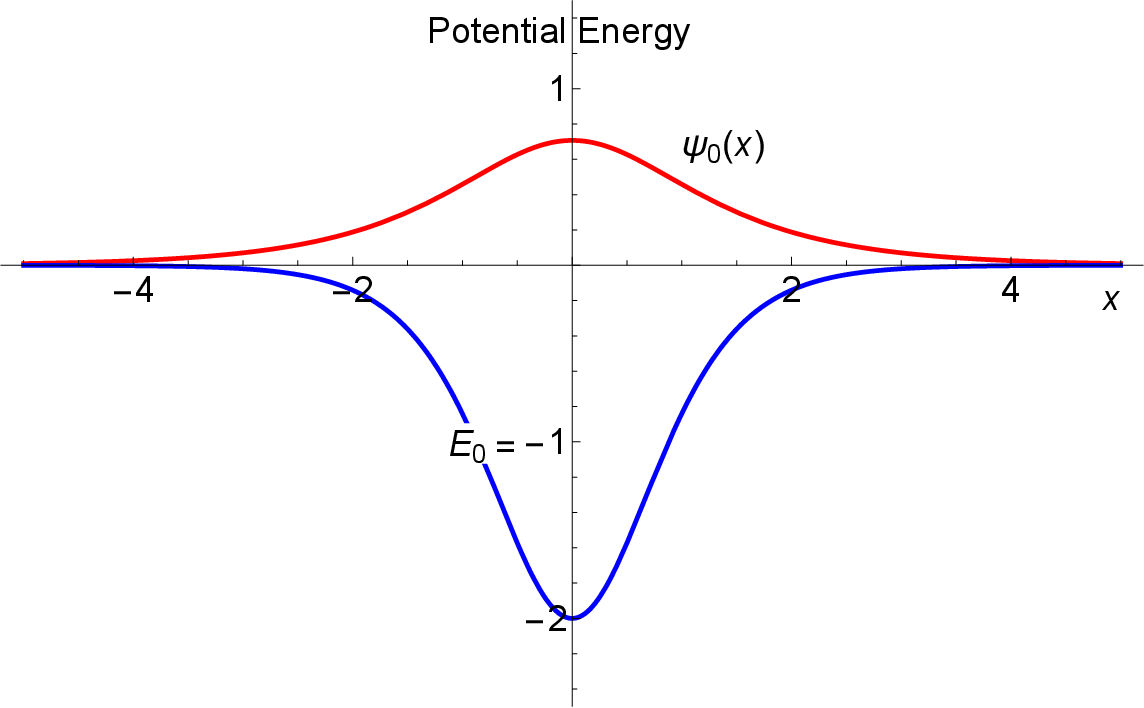}
    \caption{The reflectionless potential $V(x)=-(\hbar^2 \kappa^2/m)\sech^2 \kappa x$ (blue curve) and the bound state wave function $\psi_0(x)$ (red curve) and the corresponding energy $E_0$ is shown for $\kappa=1$ in the units $\hbar=2m=1$. Notice that the wave function $\psi_0$ has a different unit and is only shown for comparison.}  
    \label{fig:boundstates}
\end{figure}

The algebraic solution to the energy eigenstates of this problem were reviewed in an unpublished work of R. L. Jaffe \cite{Jaffe} for any integer $N$. We will restrict our attention to the case $N=1$ \cite{Kay} such that the Hamiltonian is
\begin{eqnarray}
    H= - \frac{\hbar^2}{2m} \frac{d^2}{d x^2} - \frac{\hbar^2 \kappa^2}{m \cosh^2 \kappa x} \;. \label{hamiltonian2}
\end{eqnarray}
In analogy with the harmonic oscillator problem, we define the following ``creation" and ``annihilation"  operators
\begin{eqnarray}
a^{\dagger} & := & \frac{1}{\sqrt{2m}} \left(P + i \hbar \kappa \tanh \left( \kappa X \right) \right) \;, \\
    a & := & \frac{1}{\sqrt{2m}} \left( P - i \hbar \kappa \tanh \left(\kappa X \right) \right) \;, 
\end{eqnarray}
respectively. Here $X$ and $P$ denote the standard self-adjoint position and momentum operator in the Hilbert space $L^2(\mathbb{R})$ \footnote{The operator $X$ is defined in an appropriate domain on which it is self-adjoint so that any function of $X$ can be properly defined (on perhaps a restricted domain) via the spectral theorem. From this point on, we do not deal with the technical domain issues about the operators, see the details of such issues in \cite{selfadjointXP}}. The above operators $a$ and $a^\dagger$ are equivalent to the expressions in equation (7) in Ref. \cite{CoxLekner}. By defining $A:=a^{\dagger} a$ and $B:=a a^{\dagger}$ and using the commutation relation for the position and momentum operators, we find
\begin{eqnarray}
    A & = & \frac{P^2}{2m} + \frac{\hbar^2 \kappa^2 }{2m} \mathbb{I} - \frac{\hbar^2 \kappa^2}{m} \sech^2 \left( \kappa X \right) \; \\
        B & = &  \frac{P^2}{2m} + \frac{\hbar^2 \kappa^2}{2m} \mathbb{I} \;,
\end{eqnarray}
where $\mathbb{I}$ is the identity operator. Next, we look for the ground state wave function. In analogy with the harmonic oscillator problem, the ground state, denoted by $|0\rangle$, is defined by $a |0 \rangle=0$ in the standard bra-ket notation \cite{Appel}. In coordinate representation, this equation becomes:
\begin{eqnarray}
    \left(-i \hbar \frac{d}{dx}-i \hbar \kappa \tanh \kappa x \right) \psi_0(x) = 0 \;.
\end{eqnarray}
This can be easily integrated and the general solution is given by 
\begin{eqnarray}
    \psi_0(x)= \sqrt{\frac{\kappa}{2}} \frac{1}{\cosh \kappa x} \;, \label{bswavefunction}
\end{eqnarray}
where the factor $\sqrt{\kappa/2}$ is found by the normalization condition. The graph of this bound state wave function together with the form of the potential is shown in Fig. \ref{fig:boundstates}. 
Hence, we have obtained the ground state wave function and it has no nodes, as expected. The corresponding bound state energy $E_0=-\hbar^2\kappa^2 /2m$ can be found by substituting this solution into the time-independent Schr\"{o}dinger equation.

In order to find the other part of the spectrum, we first express the operators $A$ and $B$ in terms of the free particle Hamiltonian $H_0= P^2/2m$ and our Hamiltonian $H$:
\begin{eqnarray}
    A & = & H + \frac{\hbar^2 \kappa^2}{2m} \;, \\
    B & = & H_0 + \frac{\hbar^2 \kappa^2}{2m} \;.
\end{eqnarray}
Suppose that $\psi$ is an eigenstate of the operator $A$, i.e., $A |\psi \rangle = \alpha |\psi \rangle$. That implies 
\begin{eqnarray} 
    a (A |\psi \rangle) & = &  \alpha (a |\psi \rangle)  \nonumber \\ & = & (a a^{\dagger}) a |\psi \rangle = B (a |\psi \rangle) \;.
\end{eqnarray}
We then formally obtain $B(a|\psi\rangle)= \alpha (a|\psi\rangle)$. The only exception to this statement is the ground state $|0\rangle$ since $a|0\rangle =0$ by definition. This shows that $|0\rangle$ is an eigenstate of $A$ with eigenvalue $\alpha=0$ with no corresponding eigenstate of $B$. If $|\varphi \rangle$ is an eigenstate of $B$, then it can be similarly shown that $a^{\dagger}|\varphi\rangle$ is an eigenstate of $A$ with the same eigenvalue. Then, we formally show that the Hamiltonians $H_0$ and $H$ must have the same spectrum except for the single ground state.
The above simple argument allows us to construct the eigenstates and the corresponding eigenenergies of the Hamiltonian. The idea of the construction of such a one-to-one relation between the eigenstates of the pair of Hamiltonians ($H_0$ and $H$ here) except for the ground state of the Hamiltonian $H$ is equivalent to the more general method in the context of supersymmetric quantum mechanics \cite{CooperKhareSukhatme}.

The continuum eigenstates of the free Hamiltonian $H_0$ are denoted by $|k\rangle$ and labeled by the wave number $k$. The wave function associated with this state is given by $(2\pi)^{-1/2} e^{ikx}$ and eigenenergies are $E_k=\hbar^2 k^2/2m$. Likewise, our Hamiltonian $H$ has the same continuous spectrum with the eigenenergies $E=\hbar^2 k^2/2m$. Hence, the spectrum of $H$ is constructed, that is, it consists of a single bound state for $E=-\hbar^2 \kappa^2/2m$ and a continuous spectrum on the positive real axis $E=\hbar^2 k^2/2m$ with $k$ being a (nonzero) real number. 

To construct the continuous eigenstates, let $|\tilde{k} \rangle:= a^{\dagger} |k \rangle$. Then, the action of $H$ on this state $|\tilde{k} \rangle$ becomes
\begin{eqnarray}
    H | \tilde{k} \rangle & = &  \left(A- \frac{\hbar^2 \kappa^2}{2m}\right)a^{\dagger} |k \rangle = a^{\dagger} B |k\rangle -\frac{\hbar^2 \kappa^2}{2m} a^{\dagger} |k\rangle = \frac{\hbar^2 k^2}{2m} | \tilde{k} \rangle \;.
\end{eqnarray}
Thus, $| \tilde{k} \rangle$ is a continuum eigenstate of $H$ with eigenvalue $\hbar^2 k^2/2m$ and the associated eigenfunction is given by
\begin{eqnarray}
   \phi_k(x)=  \langle x |\tilde{k} \rangle & = &  \langle x| a^{\dagger} |k\rangle = \frac{1}{\sqrt{2m}}\langle x | (P+i \hbar \kappa \tanh (\kappa X)) |k\rangle 
   =  \frac{e^{i k x}}{\sqrt{4\pi m}} \left(\hbar k + i \hbar \kappa \tanh \kappa x \right) \;,  \label{unnormalizedcontstates}
\end{eqnarray}
where $P|k\rangle = \hbar k |k\rangle$ and $\langle x|k\rangle = (2\pi)^{-1/2} e^{ikx}$. The above formula is equivalent to the equation (19) previously given in \cite{Lekner}.

The reason this potential is called reflectionless can be easily seen from the asymptotic behavior of the above continuum states and the usual scattering boundary conditions $\phi_k(x) \sim e^{ikx}+R(k) e^{-ikx}$ as $x \to -\infty$ if we send the particle from the left. Since $\phi_k(x) \sim (4\pi m)^{-1/2} e^{i k x} \left(\hbar k - i \hbar \kappa \right)$ as $x \to -\infty$, $R(k)=0$.

\section{Orthonormality and Completeness}
\label{orthonormalityandcompleteness}

When the Hamiltonian has only a discrete set of states, a generic wave function $\psi$ can be expanded in terms of the orthonormal set of eigenfunctions $\{\psi_n\}$ of the Hamiltonian operator, that is, $\psi(x)= \sum_{n} c_n \psi_n(x)$. The orthonormality relation is expressed by 
\begin{eqnarray}
 \int_{-\infty}^{\infty} \psi_{n}^{*}(x) \psi_m (x) \, d x = \delta_{nm}  \;
\end{eqnarray}
and the expansion coefficients are given by $c_n= \int_{-\infty}^{\infty} \psi_{n}^{*}(x) \psi(x) \; d x$. Here, no degeneracy label is needed for one-dimensional bound state problems for non-singular potentials. However, when we have a continuous part of the spectrum, this idea needs to be generalized. 

If the Hamiltonian under consideration has both discrete and continuous spectra, normalizable eigenfunctions and continuum eigenfunctions together  form a complete set, 
and the expansion of a wave function has the form
\begin{eqnarray}
    \psi(x)= \sum_{n} c_n \psi_n(x) +  \sum_i \int c_i(E) \psi_{i,E}(x) \; d E \;, \label{expansion}
\end{eqnarray}
where the integration is over the continuous spectrum of the Hamiltonian at hand \cite{Gasiorowicz, Konishi} \footnote{In general, there is a more general measure (such as $w(E) dE$ for some weight function). However, the form of (\ref{expansion}) would be sufficient for typical problems in quantum mechanics \cite{Galindo1}.} and the index $i$ refers to a possible degeneracy of continuum states. The completeness relation in this general case is given by 
\begin{eqnarray}
        \sum_n \psi_{n}^{*}(x) \psi_n(y) +  \sum_i \int (\psi_{i,E}(x))^{*} \psi_{i, E}(y) \; d E = \delta(x-y) \;.
\end{eqnarray}
Here we consider the case where the discrete spectrum does not overlap with the continuous spectrum of $H$. The orthonormality relation for the continuum states is given by
\begin{eqnarray}
    \int_{-\infty}^{\infty} (\psi_{i, k_1}(x))^{*} \psi_{j, k_2}(x) \; d x = \delta_{ij} \delta (k_1-k_2) \;. \label{orth2}
\end{eqnarray}
It is important to notice that we have expressed the orthogonality relation for the continuum eigenfunctions in terms of the wave number $k$, where $E=\hbar^2 k^2/2m$ for convenience. The completeness relation for the entire set of eigenfunctions $\{\psi_n, \psi_{i, k} \}$ is then given by 
\begin{eqnarray}
    \sum_{n} \psi_{n}^{*}(x) \psi_{n}(y) + \sum_i \int_{-\infty}^{\infty} (\psi_{i, k}(x))^{*} \psi_{i, k}(y) \; d k = \delta(x-y) \;.  \label{completeness}
\end{eqnarray}
The orthonormality and the completeness relations are general properties of self-adjoint operators in Hilbert spaces. However, it would be useful from the pedagogical point of view if one could demonstrate these properties explicitly for the eigenfunctions of some exactly solvable Hamiltonians. We  now demonstrate the completeness relation for the well-known reflectionless potential discussed in many elementary quantum textbooks \cite{Griffiths}.

\section{Orthonormality and Completeness Relations for the Reflectionless Potential}
\label{Completeness Relation for the Reflectionless Potential}

The normalization constants of the continuum states contains important information about the system and the necessary step for the correct completeness relation \cite{Lin}. It can be easily found using the Dirac's bra-ket notation:
\begin{eqnarray}
 \int_{-\infty}^{\infty} \phi_{k_1}^{*}(x) \phi_{k_2}(x) \, dx & = &  \langle \tilde{k}_1|\tilde{k}_2 \rangle  = \langle k_1 | a a^{\dagger} |k_2 \rangle  \nonumber \\ & = & \langle k_1 | \left(H_0 + \frac{\hbar^2 \kappa^2}{2m}\right) |k_2 \rangle \nonumber \\ & = &  \frac{\hbar^2}{2m}(k_{2}^2 + \kappa^2) \langle k_1  |k_2 \rangle \nonumber \\ & = & \frac{\hbar^2}{2m} (k_{2}^2 + \kappa^2) \delta(k_1-k_2) \;. \label{normalization2}
\end{eqnarray}
One can also derive the above orthogonality relation using the explicit form of the continuum eigenstates (\ref{unnormalizedcontstates}), as shown in Appendix. From the above result (\ref{normalization2}), we choose the following normalized continuum eigenfunctions
\begin{eqnarray}
    \psi_k(x) = \langle x |\psi_k \rangle = \frac{\phi_k(x)}{\frac{1}{\sqrt{2m}}\left(\hbar \kappa + i \hbar k \right)} = \frac{e^{i k x}}{\sqrt{2\pi}} \frac{\left(k + i \kappa \tanh \kappa x \right) }{\kappa + i k} \;. \label{contstates}
\end{eqnarray}
Using the normalized continuum states (\ref{contstates}), we obtain 

\begin{eqnarray}
\int_{-\infty}^{\infty} \psi_{k}^{*}(x) \psi_k(y) \, d k = \int_{-\infty}^{\infty} \frac{e^{ik (y-x)}}{k^2 + \kappa^2} \Bigg(k^2 + i k \kappa\left(\tanh \kappa y- \tanh \kappa x\right) + \kappa^2 \tanh \kappa x \tanh \kappa y \Bigg) \frac{dk}{2\pi} \;. \label{completeness1}
\end{eqnarray}
Adding and subtracting $\kappa^2$ inside the bigger bracket, the above integral can be written as 
\begin{eqnarray} & & 
    \int_{-\infty}^{\infty} e^{ik (y-x)} \frac{dk}{2\pi} -\kappa^2 (1- \tanh \kappa x \tanh \kappa y ) \int_{-\infty}^{\infty} \frac{e^{ik(y-x)}}{k^2 + \kappa^2}
\, \frac{dk}{2\pi} \nonumber \\ & & \hspace{5cm} +    
    \kappa\left(\tanh \kappa y- \tanh \kappa x\right) \int_{-\infty}^{\infty}  \frac{i k  e^{ik(y-x)}}{k^2 + \kappa^2}
\, \frac{dk}{2\pi} \;. \label{completenesspart1}
\end{eqnarray}
The first term is just the integral representation of the Dirac delta function $\delta(x-y)$. The second integral arises typically when dealing with  Green's functions, propagators, and one-dimensional potential scattering problems and can be easily evaluated by choosing a contour consisting of a finite interval along the real $k$ axis and closing it by upper (lower) semi-circle in the complex $k$ plane if $y-x>0$ ($y-x<0$). Since there are two simple poles at $\pm i \kappa$ the integral can be found by applying the residue theorem in complex analysis:
\begin{eqnarray}
    \int_{-\infty}^{\infty} \frac{e^{ik(y-x)}}{k^2 + \kappa^2}
\, \frac{dk}{2\pi} = \frac{1}{2\kappa} \, e^{-\kappa|y-x|} \;.
\end{eqnarray}
The last integral can also be evaluated by the method of differentiation under the integral sign as
\begin{eqnarray}
    \int_{-\infty}^{\infty} \frac{i k e^{ik(y-x)}}{k^2 + \kappa^2}
\, \frac{dk}{2\pi} & = &  - \frac{\partial}{\partial x}  \int_{-\infty}^{\infty} \frac{e^{ik(y-x)}}{k^2 + \kappa^2} \, \frac{dk}{2\pi} \nonumber \\ & = &  - \frac{\partial}{\partial x} 
\left( \frac{1}{2\kappa} \, e^{-\kappa|y-x|} \right) \nonumber \\ & = & - \frac{\sgn(y-x)}{2} e^{-\kappa|y-x|} \;,
\end{eqnarray}
where $\sgn(y-x)$ is the sign function. Combining all these integrals in (\ref{completenesspart1}), and using the addition formulas for hyperbolic functions, the right-hand side of the equation (\ref{completeness1}) simplifies to
\begin{eqnarray}
\delta(x-y)-\frac{\kappa}{2} \frac{e^{-\kappa|y-x|}}{\cosh \kappa x \cosh \kappa y} \left( \cosh \kappa (y-x) + \sgn(y-x) \sinh \kappa(y-x)\right) \;.
\end{eqnarray}
Then, using the identity 
\begin{eqnarray}
    \cosh \kappa (y-x) + \sgn(y-x) \sinh \kappa(y-x) = \cosh \kappa |y-x| + \sinh \kappa|y-x| = e^{\kappa|y-x|} \;,
\end{eqnarray}
we finally obtain
\begin{eqnarray}
    \int_{-\infty}^{\infty} \psi_{k}^{*}(x) \psi_k(y) \, d k =  \delta(x-y)  -\frac{\kappa}{2} \frac{1}{\cosh \kappa x \cosh \kappa y} \;.
\end{eqnarray}
This shows that continuum eigenstates are not sufficient for the completeness relation, we must also take the presence of the
bound states into account.

The present model has double degeneracy in the continuum due to the parity invariance. For this reason, one can introduce the even- and odd-parity solutions and prove the completeness relation using these states. 
This is accomplished in the following section.

\section{Completeness Relation Using the Even and Odd Parity Eigenstates}

Since our Hamiltonian is invariant under the parity operator $\Pi$,  we can decompose our Hilbert space into a direct sum of parity eigenstates. Recall that $(\Pi \psi)(x)=\psi(-x)$, for any function $\psi$, and this is a unitary transformation in the Hilbert space as the norm of a function does not change. Moreover, $\Pi^2=\mathbb{I}$, and $\Pi^\dag=\Pi$, so that the parity operator has eigenvalues $\pm 1$. Here $\mathbb{I}$ is the identity operator. One can check that $[\Pi, H]=0$, as our potential is an even function of $x$. Hence, we can decompose the Hilbert space into even and odd parity states and {\it then diagonalize} our Hamiltonian.  
The even and odd parity eigenfunctions for the continuum states using the expression in (\ref{contstates}) are
\begin{eqnarray}
    \psi_{k}^{e} (x) & = & \frac{1}{\sqrt{2}}(\mathbb{I}+\Pi) \psi_k (x) = \frac{1}{\sqrt{2}} (\psi_k(x)+\psi_k(-x)) = \frac{1}{\sqrt{\pi}} \left( \frac{k \cos kx}{\kappa+i k} - \frac{\kappa \sin k x \tanh \kappa x}{\kappa + i k}\right)  \;, \label{evenparityeigenstate} \\   
    \psi_{k}^{o} (x) & = & \frac{1}{\sqrt{2}}(\mathbb{I}- \Pi) \psi_k (x) = \frac{1}{\sqrt{2}} (\psi_k(x)-\psi_k(-x)) = \frac{i}{\sqrt{\pi}} \left(\frac{k \sin kx}{\kappa+i k} + \frac{\kappa \cos k x \tanh \kappa x}{\kappa + i k}\right) \;, \label{oddparityeigenstate}
\end{eqnarray}
and they are simultaneous eigenstates of Hamiltonian $H$ and parity $\Pi$ and equivalent to equations (11) and (14) in \cite{Lekner}. The momentum operator transforms as $P\mapsto -P$ under parity operation $\Pi$, that is, $\Pi P \Pi = - P$ (or equivalently $\Pi$ and $P$ anti-commutes \cite{Shankar}). When we consider the free Hamiltonian,  this has the effect of switching momentum eigenvalues $k$ to $-k$. This means that one can obtain negative momentum states by applying parity operation. The even parity subspace is mapped by the momentum operator to the odd parity one and vice versa since $(\mathbb{I}+\Pi)P=P  (\mathbb{I}-\Pi)$, and similarly, diagonal projections are zero. Hence the momentum operator has the form
\begin{eqnarray}
    \begin{pmatrix}
0 & P_{eo}\\
P_{oe} & 0
\end{pmatrix} \;,
\end{eqnarray}
where $P_{eo}$ is the matrix element of $P$ between the even parity and odd parity states, and similarly for $P_{oe}$.

Indeed a simple calculation, sketched below,  would show that the application of $P$ on the even/odd parity eigenstates of our Hamiltonian, gives $k$ times almost the other state but not exactly. For instance, when $P$ acts on even parity states, it also generates an additional square integrable odd wave function. This square integrable part is intimately connected to the potential, as $P$ does not commute with the Hamiltonian, it cannot preserve the eigenstates (continuous part of the spectrum). To see this, consider the following argument. If we operate $P$ on the even subspace spanned by the normalized vectors $|\psi_{k}^{e} \rangle = 2^{-1/2}(\mathbb{I}+\Pi) |\psi_k \rangle$, we see that 
\begin{eqnarray}
P |\psi_{k}^{e} \rangle = P{1\over \sqrt{2}}(\mathbb{I}+\Pi) \frac{\sqrt{2m}}{\hbar \kappa + i \hbar k}|\tilde k\rangle= {1\over\sqrt{2}} \frac{\sqrt{2m}}{\hbar \kappa + i \hbar k} (\mathbb{I}-\Pi) \left(a^\dagger+  {\hbar \kappa^2\over \sqrt{2m} k \cosh^2 (\kappa X)} \right) P|k\rangle \;,
\end{eqnarray}
where we have used the anti-commutativity of the momentum operator with the parity operator and the fact that $[P, a^\dagger]= \hbar^2 \kappa^2/(\sqrt{2m}\cosh^2 (\kappa X))$. Then,
\begin{eqnarray} \langle x| P |\psi_{k}^{e} \rangle & = & \Big({\hbar k \over \sqrt{\pi}}{\sin(kx)\over\kappa+ik} +{\hbar \over \sqrt{\pi}} \frac{\kappa \cos k x \tanh \kappa x}{\kappa + i k}\Big)+ { 1 \over \sqrt{\pi}} {i \hbar \kappa^2 \over \kappa +ik} {\sin(k x) \over \cosh^2 (\kappa x)} \nonumber \\ & = & \hbar k \psi_{k}^{o}(x) +  { 1\over \sqrt{\pi}} {i \hbar \kappa^2 \over \kappa +ik} {\sin(k x) \over \cosh^2 (\kappa x)} \;,
\end{eqnarray}
which shows how the extra term appears and makes $P$ {\it almost} skew-diagonal. Note that since $\psi^{o}_k(x)$ is a basis of odd functions of  $L^2({\mathbb{R}})$, we can expand the additional function in this odd basis giving us off-diagonal elements. To see how this term behaves, consider the matrix elements
\begin{eqnarray}
    \langle \psi_{k'}^{o} |P | \psi_{k}^{e} \rangle = \int_{-\infty}^{\infty} \left(\psi_{k'}^{o}(x)\right)^{*} \langle x|P| \psi_{k}^{e} \rangle \; d x \;.
\end{eqnarray}
Substituting the above result for $\langle x|P| \psi_{k}^{e} \rangle$ and the odd-parity eigenstates $\psi_{k'}^{o}$, we obtain
\begin{eqnarray}
     & & \langle \psi_{k'}^{o} |P | \psi_{k}^{e} \rangle = \hbar k \delta(k-k') + \frac{i \hbar \kappa^2}{ \pi (\kappa+ik)(\kappa-ik')}\int_{-\infty}^{\infty} 
     \frac{\sin k' x \sin k x}{\cosh^2 \kappa x} \; d x  \nonumber \\ & & \hspace{5cm} + \frac{i \hbar \kappa^3}{ \pi k' (\kappa+ik)(\kappa-ik')}\int_{-\infty}^{\infty} 
     \frac{\cos k' x \sin k x \tanh \kappa x}{\cosh^2 \kappa x} \; d x  \;,
\end{eqnarray}
where we have used the orthonormality of the odd-parity eigenstates. Using the fact that $d/dx(\sech^2 \kappa x) = -2\kappa \sech^2 \kappa x \tanh \kappa x$ and the integration by parts in the second integral, we get
\begin{eqnarray}
     & & \langle \psi_{k'}^{o} |P | \psi_{k}^{e} \rangle = \hbar k \delta(k-k') + \frac{i \hbar \kappa^2}{ \pi (\kappa+ik)(\kappa-ik')}\int_{-\infty}^{\infty} 
     \sech^2 \kappa x \sin k' x \sin k x \; d x  \nonumber \\ & & \hspace{2cm} + \frac{i \hbar \kappa^3}{2\pi k' (\kappa+ik)(\kappa-ik')} \int_{-\infty}^{\infty} \sech^2 \kappa x \left(
     k \cos k' x \cos k x -k' \sin k'x \sin k x \right)\; d x  \;.
\end{eqnarray}
From the trigonometric identities $\sin k' x \sin k x= (1/2)\left(\cos (k-k')x - \cos (k+k')x \right)$ and $\cos k' x \cos k x= (1/2)\left(\cos (k-k')x + \cos (k+k')x \right)$, the above integrals are the Fourier transform of $\sech^2 \kappa x$. This has been computed in Appendix and given by the equations (\ref{ftsech1}) and (\ref{I2}). Hence we obtain
\begin{eqnarray}
     & & \langle \psi_{k'}^{o} |P | \psi_{k}^{e} \rangle = \hbar k \delta(k-k') + \frac{i \hbar \kappa}{ 2\pi (\kappa+ik)(\kappa-ik')} \left(\frac{(k-k')}{\sinh \frac{\pi(k-k')}{2\kappa}} - \frac{(k+k')}{\sinh \frac{\pi(k+k')}{2\kappa}} \right) \nonumber \\ & & \hspace{4cm} + \frac{i \hbar \kappa^2}{4 \pi k' (\kappa+ik)(\kappa-ik')} \left(\frac{(k-k')^2}{\sinh \frac{\pi(k-k')}{2\kappa}} + \frac{(k+k')^2}{\sinh \frac{\pi(k+k')}{2\kappa}} \right) \;.
\end{eqnarray}
This expression explicitly shows that there is a skew-diagonal leading part (which is unbounded as $k$ grows) and  the additional matrix elements are actually decaying rapidly with $k,k'$ (when $k$ and $k'$ are not equal). 

We now compute the second term in the completeness relation (\ref{completeness}) using the above common eigenstates of Hamiltonian and parity operator and restricting the integral over $k>0$. The second term consists of two parts, namely even ($i=e$) and odd parity ($i=o$) parts. It follows that
\begin{eqnarray}
    & & \hskip-2cm \int_{0}^{\infty} (\psi_{k}^{e}(x))^{*} \psi_{k}^{e}(y) \; d k +  \int_{0}^{\infty} (\psi_{k}^{o}(x))^{*} \psi_{k}^{o}(y) \; d k \nonumber \\ 
    & & = \frac{1}{\pi} \Bigg( \int_{0}^{\infty} \cos k (x-y) \; d k + \kappa^2 (\tanh \kappa x \tanh \kappa y - 1) \int_{0}^{\infty} \frac{\cos k(x-y)}{\kappa^2 + k^2} \; dk \nonumber \\ & & \hspace{2cm} +  \kappa (\tanh \kappa y - \tanh \kappa x) \int_{0}^{\infty} \frac{k \sin k(x-y)}{\kappa^2 + k^2} \; dk  \Bigg) \;.
\end{eqnarray}
The first integral should be understood in the distributional sense and one can find it by expressing the cosine function in terms of the complex exponential function
$\int_{0}^{\infty} \cos k(x-y) \, dk = (1/2) \int_{-\infty}^{\infty} \cos k(x-y) \, dk = (1/2) \Real(\int_{-\infty}^{\infty} \exp i k(x-y) \, dk)=\pi \delta(x-y)$. The other two integrals can be evaluated easily by first replacing the integrals over the half line with the half of the integral over the entire real axis, then writing the sine or cosine function in the complex exponential form, and then choosing a semi-circular contour in the complex plane for each such term in the appropriate upper of lower half-plane.  The simple poles are located at $k= \pm i \kappa$. Therefore, we obtain
\begin{eqnarray}
    & & \hskip-2cm \int_{0}^{\infty} (\psi_{k}^{e}(x))^{*} \psi_{k}^{e}(y) \; d k +  \int_{0}^{\infty} (\psi_{k}^{o}(x))^{*} \psi_{k}^{o}(y) \; d k \nonumber \\ 
    & & = \delta(x-y) + \frac{\kappa}{2} (\tanh \kappa x \tanh \kappa y - 1) \exp(-\kappa|x-y|) \nonumber \\ & & \hspace{2cm} +  \frac{\kappa}{2}(\tanh \kappa y - \tanh \kappa x) \sgn(x-y) \exp(-\kappa|x-y|) \;.
\end{eqnarray}
Hence, after simplifications, we obtain
\begin{eqnarray}
    \int_{0}^{\infty} (\psi_{k}^{e}(x))^{*} \psi_{k}^{e}(y) \; d k +  \int_{0}^{\infty} (\psi_{k}^{o}(x))^{*} \psi_{k}^{o}(y) \; d k 
     =  \delta(x-y) - \frac{\kappa}{2 \cosh \kappa x \cosh \kappa y}    \;. \label{contrcont}
\end{eqnarray}

\section{Results}

If we substitute the contribution of the continuum states given above (\ref{contrcont}) into the general completeness relation (\ref{completeness}), we have
\begin{eqnarray}
         \sum_{n} \psi_{n}^{*}(x) \psi_{n}(y) = \frac{\kappa}{2 \cosh \kappa x \cosh \kappa y}   \;. \label{completenessfinal}
\end{eqnarray}
The number of bound states can be easily found by the following nice argument. Set $y=x$ and integrate both sides with respect to $x$ over the entire real axis. Since all $\psi_{n}(x)$ are assumed to be normalized, the left-hand side of the above expression yields the number of bound states $N$. Using the result $\int_{-\infty}^{\infty} \sech^2 \kappa x \, dx  = 2/\kappa$ (this can be obtained as the limiting case of the result (\ref{ftsech1}) in Appendix as $k \to 0^+$), we obtain $\sum_{n=0} \int_{-\infty}^{\infty} |\psi_n(x)|^2 d x =1$. Since each bound state is normalized, there can only be a single bound state ($n=0$) for this system. It follows from the relation (\ref{completenessfinal}) that the bound state wave function can be uniquely determined and given by $\psi_0(x)= \sqrt{\kappa/2} \sech \kappa x$ with the correct normalization factor, which is exactly (\ref{bswavefunction}). Finding the number of bound states and the bound state wave function from the knowledge of continuum states can be considered as the simpler version of inverse problem for bound states, which can be described as finding the potential and wave functions from the knowledge of a complete set of negative bound state energies and the reflection and transmission coefficients for all positive energies \cite{Gutierrez}. This is referred as the Mar\u{c}henko method in the literature \cite{Schumayer}, which is a special case of Gel'fand Levitan equation \cite{Novikov}. An extension of the Mar\u{c}henko method developed by Kay and Moses \cite{Kay} has been also used to extract the associated bound state and scattering wave functions from a given finite number of energy eigenvalues.



\section{Appendix}

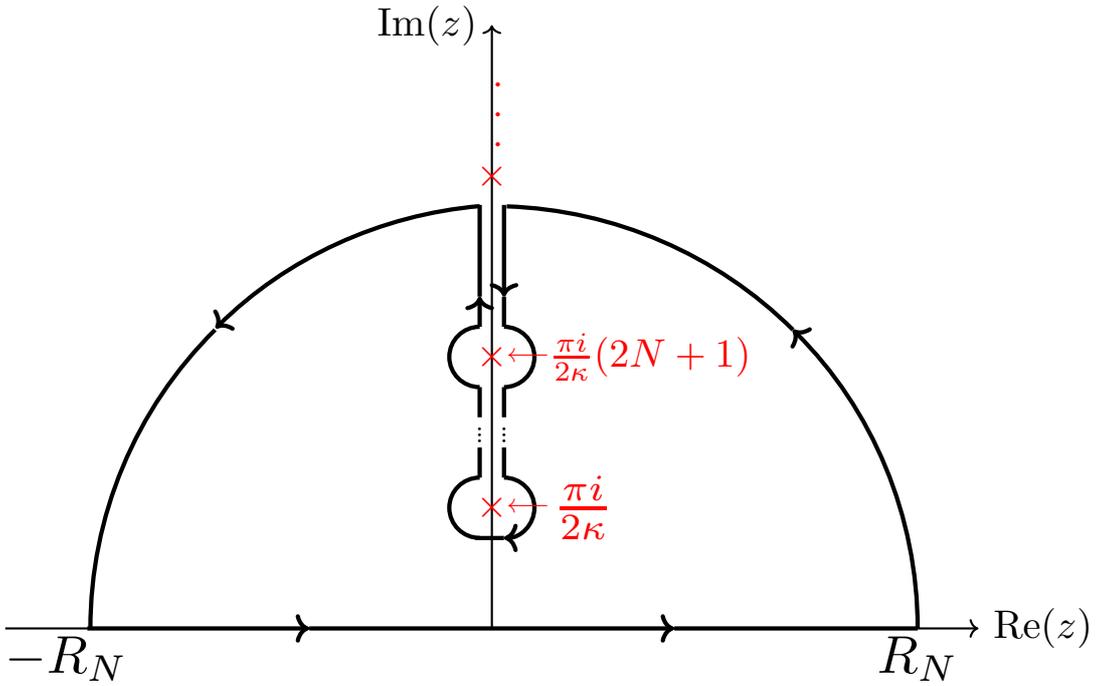
\begin{figure}
    \begin{center}
\begin{tikzpicture}[scale=0.8,
decoration={
    markings,
    mark=between positions 1.2 and 1.8 step 0.2 with {\arrowreversed{stealth}}}
                ] 
        \draw[ultra thick, black, ->] (7,0) arc [start angle=0, end angle=45,
		x radius=7cm, y radius=7cm];
        \draw[ultra thick, black] (4.95, 4.95) arc [start angle=45, end angle=88,
		x radius=7cm, y radius=7cm];
		
		\draw[ultra thick, black, ->] (-0.2,7) arc [start angle=95, end angle=135,
		x radius=7cm, y radius=7cm];
        \draw[ultra thick, black] (-4.55, 4.95) arc [start angle=135, end angle=180,
		x radius=7cm, y radius=7cm];

        \draw[ultra thick, ->] (0.2,7.02) -- (0.2,5.5);
         \draw[ultra thick] (0.2,5.5) -- (0.2,5);
         \draw[ultra thick] (-0.2,7.02) -- (-0.2,5.5);
         \draw[ultra thick, <-] (-0.2,5.5) -- (-0.2,5);

        \draw[ultra thick, black] (0.2,5) arc [start angle=90, end angle=-90,
		x radius=0.5cm, y radius=0.5cm];

        \draw[ultra thick] (0.2,4) -- (0.2,3.5);

        \node at (0.2, 3.3) {$\cdot$};
        \node at (0.2, 3.2) {$\cdot$};
        \node at (0.2, 3.1) {$\cdot$};

        \draw[ultra thick] (0.2,3) -- (0.2,2.5);

        \draw[ultra thick, black] (-0.2,1.5) arc [start angle=-90, end angle=-270,
		x radius=0.5cm, y radius=0.5cm];

        \draw[ultra thick, black, ->] (0.2,2.5) arc [start angle=90, end angle=-90,
		x radius=0.5cm, y radius=0.5cm];

         \draw[ultra thick, black] (0.2,1.5) arc [start angle=0, end angle=-100,
		x radius=0.4cm, y radius=0.002cm];

        \draw[ultra thick] (-0.2,2.5) -- (-0.2,3);

        \node at (-0.2, 3.3) {$\cdot$};
        \node at (-0.2, 3.2) {$\cdot$};
        \node at (-0.2, 3.1) {$\cdot$};

        \draw[ultra thick] (-0.2,3.5) -- (-0.2,4);

        \draw[ultra thick, black] (-0.2,4) arc [start angle=-90, end angle=-270,
		x radius=0.5cm, y radius=0.5cm];

        \draw[thick] (-8,0) -- (-4,0);
		\draw[thick] (-4,0) -- (4,0);
  		\draw[thick,->] (4,0) -- (8,0) node[right, scale=1.5] {$\Re(z)$};
		\draw[thick, ->] (0,0) -- (0,10) node[left, scale=1.5] {$\Im(z)$};
        \draw[ultra thick, ->] (-6.65,0) -- (-3,0);
         \draw[ultra thick, ->] (-3,0) -- (3,0);
          \draw[ultra thick] (3,0) -- (7,0);
		\node [scale=1.5] at (0,2) {${\color{red}\times}$}; 
        \node [scale=2] at (1.5,2) {${\color{red}\frac{\pi i}{2 \kappa}}$};	
         \node at (0.6,2) {${\color{red}\longleftarrow}$};	
        \node [scale=1.5] at (0,4.5) {${\color{red}\times}$}; 
        \node [scale=1.5] at (2.6,4.5) {${\color{red}\frac{\pi i}{2 \kappa}(2N+1)}$};	
        \node at (0.6,4.5) {${\color{red}\longleftarrow}$};	
        \node [scale=1.5] at (0,7.5) {${\color{red}\times}$};
         \node [scale=1.5] at (0.1, 8) {${\color{red}\cdot}$};
        \node [scale=1.5] at (0.1, 8.5) {${\color{red}\cdot}$};
        \node [scale=1.5] at (0.1, 9) {${\color{red}\cdot}$};
       \node [scale=2] at (7,-0.5) {$R_N$};
       \node [scale=2] at (-7,-0.5) {$-R_N$};
 \end{tikzpicture}     
\end{center}
\caption{The choice of the closed contour $C$ for the integral $I_1$.}
    \label{fig:contour}
\end{figure}

We can alternatively normalize the continuum eigenstate by substituting its explicit form (2.12) into the orthonormality relation (3.4):
\begin{eqnarray}
     \int_{-\infty}^{\infty} (\phi_{k_1}(x))^{*} \phi_{k_2} (x) \; d x & = & \frac{\hbar^2}{4\pi m} \Bigg( k_1 k_2 \int_{-\infty}^{\infty} e^{i(k_2-k_1)x} d x 
     + i \kappa (k_1-k_2) \int_{-\infty}^{\infty} e^{i(k_2-k_1)x} \tanh \kappa x \; d x \nonumber \\ & + & \kappa^2 \int_{-\infty}^{\infty} e^{i(k_2-k_1)x} \tanh^2 \kappa x \; d x  \Bigg)  \;.
\end{eqnarray}
The expression $\int_{-\infty}^{\infty} e^{i(k_2-k_1)x} d x$ is equal to $2\pi \delta(k_2-k_1)$. Using the identity $\tanh^2 \kappa x = 1- \sech^2 \kappa x$ in the last integral, we get one more Dirac delta function, and the other terms include following the integrals 
\begin{eqnarray}
    I_1:= \int_{-\infty}^{\infty} \sech^2 \kappa x \, \exp(i k x) \; dx \;,  \hspace{1cm}  I_2:= \int_{-\infty}^{\infty} \tanh \kappa x \, \exp(i k x) \; dx \;.
\end{eqnarray}
For $I_1$, $\sech^2 \kappa x$ has simple poles located at the imaginary axis $z=z_n= \pm (\pi i /2 \kappa) (2n+1)$ for $n=0, 1, 2, \ldots$. When $k>0$, we choose the closed contour $C$ (see Fig. \ref{fig:contour}) consisting of the straight line segment along the real axis from $-R_N$ to $R_N$ such that $|z_N|< R_N < |z_{N+1}|$ and the deformed semi-circular path $C_{R_N}$ of radius $R_N$ in the upper half plane such that isolated poles up to $z_N$ in the imaginary axis are all removed. 

In this case, the residue theorem yields   
\begin{eqnarray}
    \int_{-R_N}^{R_N}  \sech^2 \kappa x \, e^{i k x} \; dx + \int_{C_{R_N}}  \sech^2 \kappa z \, e^{i k z} \; dz - 2 \pi i \sum_{j=1}^{N} \res{\left( \sech^2 \kappa z \; e^{i k z}; z_n \right)} =0 \;,
\end{eqnarray}
where the last term is the result of the coinciding limit of the vertical lines along the imaginary axis so that we end up with integrals over clockwise oriented $N$ small circles. After taking the limit of the above equation as $N \to \infty$ (which is equivalent to $R_N \to \infty$) and applying Jordan's lemma, the second integral goes to zero so that we find
\begin{eqnarray}
    I_1 = \frac{2\pi i k}{i \kappa^2}  \sum_{n=0}^{\infty} \exp\left(-\frac{\pi k}{2 \kappa}(2n+1)\right) = \frac{\pi k}{\kappa^2} \frac{1}{\sinh \frac{\pi k}{2 \kappa}} \;,
\end{eqnarray}
for $k>0$. The result of the integral $I_1$ is also given in 3.982 (1) in \cite{GradRyznik}. The case for $k<0$ is equivalent to the integral by mapping $x \to -x$ in the above result. In this case, we obtain the same formula so we find the desired result $I_1$ for $k<0$. As a corollary of this result $\int_{-\infty}^{\infty} \sech^2 \kappa x \, dx  = \lim_{k\to 0} I_1 = 2/\kappa$.

The integral $I_2$ is the complex conjugate of the Fourier transform of the hyperbolic tangent function. However, it can not be evaluated directly by the above contour integration technique since Jordan's lemma does not apply ($\tanh \kappa z$ does not decay at infinity) and the integral does not converge in the ordinary sense. Nevertheless, the integral can be still understood in the principal value sense and can be evaluated by expressing $\sech^2 \kappa x= \frac{1}{\kappa} \frac{d}{dx} \tanh \kappa x$ in the integral $I_1$ and applying the fact that the Fourier transform of the derivative of a function is equal to the product of the function with the Fourier variable $k$ \cite{Appel}, i.e., 
\begin{eqnarray}
  I_1 = \frac{1}{\kappa} \mathcal{F} \left((\tanh \kappa x)' \right) (k) = -i k \mathcal{F}(\tanh \kappa x)(k) = -i k I_2 \;,  \label{ftsech1}
\end{eqnarray}
where $'$ denotes the derivative with respect to $x$ and $\mathcal{F}(f)(k):=\int_{-\infty}^{\infty} f(x) e^{-ikx} dx$. Then, from this relation, we obtain 
\begin{eqnarray}
    I_2= \frac{\pi i}{\kappa} \frac{1}{\sinh \frac{\pi k}{2 \kappa}} \;. \label{I2} 
\end{eqnarray}
Hence, by combining all these results, we obtain
\begin{eqnarray}
         \int_{-\infty}^{\infty} \phi_{k_1}^{*}(x) \phi_{k_2} (x) \; d x = \frac{\hbar^2}{2m} (k_{1}^{2} + \kappa^2) \delta(k_1-k_2) = \frac{\hbar^2}{2m} (k_{2}^{2} + \kappa^2) \delta(k_1-k_2) \;,
\end{eqnarray}
from which we can normalize the continuum eigenstates as $\psi_k(x)= \sqrt{2m} \phi_k(x)/\hbar \kappa+i \hbar k$.

\section*{Acknowledgements}
We would like to express our gratitude to the anonymous reviewers for their exceptional guidance and help. Indeed, their suggestions and corrections restructured our original manuscript, their kind efforts reshaped our modest draft into a coherent and enjoyable piece of work that we hope the readers do appreciate. O. T. Turgut would like to thank Prof. M. Deserno and Prof. S. Dodelson for an enjoyable visit to Carnegie-Mellon University while this work is being completed.

\end{document}